\documentclass[a4paper,12pt]{article}

\usepackage[english]{babel}
\usepackage{amsmath,amsfonts,amssymb}
\usepackage{hyperref}

\def\eps{\varepsilon}
\def\mult{\,}

\begin{document}
\title{Calculation of the dynamical critical exponent in the model~$A$ of critical dynamics to order $\eps^4$}
\author{
  L. Ts. Adzhemyan\footnote{Department of Theoretical Physics, St.~Petersburg
  University, Ulyanovskaya 1, St.~Petersburg, Petrodvorets, 198504,
  Russia. E-mail: \tt Loran.Adzhemyan@pobox.spbu.ru, leo\_ff@mail.ru},
  \ S. V. Novikov\footnote{Google St.~Petersburg, Alia Tempora, ul.
  Mayakovskogo, Bldg 3B, Floors 8-9, St.~Petersburg, 191025,
  Russia. E-mail: \tt sergeyn@google.com},
  \ L. Sladkoff${}^*$
 }
\date{}

\maketitle
\begin{abstract}
A new method based on the $R'$-operation of the renormalization
theory is proposed for the numerical calculation of the
renormalization constants in the theory of critical behaviour. The
problem of finding residues of the poles of the Green's functions
at $\eps=0$, where $\eps=4-d$, is reduced to the evaluation of
multiple UV-finite integrals, which can be performed by means of
standard integration programs. The method is used to calculate the
renormalization group functions of the model $A$ of critical
dynamics in four-loop approximation. Dynamical exponent $z$ of the
model $A$ is calculated in the fourth order of the
$\eps$-expansion.
\end{abstract}

\section*{Introduction}
The method of the renormalization group (RG) nowadays is the main
tool of calculation of critical exponents in the theory of
critical phenomena. The basis of that method is the technique of
the ultraviolet renormalization of a model. Critical exponents are
found from the calculated renormalization constants by means of
the standard RG rules. Thus calculation of the renormalization
constants is the primary concern of the method. The most
consistent scheme of applying the RG method is combining it with
the $\eps$-expansion (with $\eps=4-d$, where $d$ is spatial
dimension) and that is the way it is used throughout the paper.

Creation of subtle analytical methods of calculations allowed to
attain 5-loop accuracy (fifth order of the $\eps$-expansion) in
the theory of static critical phenomena~\cite{5loop} and maximum
3-loop accuracy (third order of the $\eps$-expansion) in the
theory of critical dynamics~\cite{3loop}. These record results
have held since 1991 and 1984 respectively. Attempts of analytical
calculations in higher orders met fundamental
difficulties~\cite{KS}.

For numerical calculation of the renormalization constants it is
essential to be able to extract residues at $\eps=0$ from the
graphs. We developed such technique based on well-known
$R'$-operation of the theory of renormalizations. It reduces
calculation of renormalization constants to the evaluation of
multiple UV-convergent integrals which can be performed by
standard programs for numerical integration. We used the technique
for calculation of dynamical exponent of the model A in 4-loop
approximation (fourth order of the $\eps$-expansion). Brief
formulation of the model and the results obtained are given below.

\section{Model}
The model A describes dynamical behavior of pure homogeneous Ising
system like uniaxial magnetic materials. Particularly the critical
dynamical exponent $z$ determines the growth of the relaxation
time near the critical point. This paper presents the results of
4-loop calculation of the exponent $z$ with the help of the
renormalization group (RG) in the $\eps=4-d$ expansion framework.

The model A in the field-theoretic formulation (see e.g.~\cite{V})
can be described by the action
\begin{equation}
S(\psi, \psi') = \lambda_0 \psi' \psi' + \psi' \left[ -\partial_t \psi +
 \lambda_0 ( \partial^2 \psi - \tau_0 \psi - g_0 \psi^3 / 6 )
 \right],
\label{initAction}
\end{equation}
with the $n$-component order parameter field $\psi({\bf x}, t)$
and the auxiliary field $\psi'({\bf x}, t)$. At dimension
$d=4-\eps$ graphs of the perturbation theory are ultraviolet
divergent that manifest itself as pole at $\eps=0$. These
divergences are eliminated by introducing the renormalization
constants $Z$ for which we use the minimal subtraction scheme
(MS). Renormalized action has the form
\begin{equation}
S_R(\psi, \psi') = Z_1 \lambda \psi' \psi' + \psi' \left[ -Z_2
\partial_t \psi +
 \lambda ( Z_3 \partial^2 \psi - Z_4 \tau \psi - Z_5 g \mu^\eps \psi^3 / 6 )
 \right].
\label{renormAction}
\end{equation}
It is obtained from (\ref{initAction}) by multiplicative
renormalization of the parameters $\lambda_0=\lambda Z_\lambda$,
$\tau_0=\tau Z_\tau$, $g_0=g \mu^\eps Z_g$ and the fields $\psi
\to Z_\psi \psi$, $\psi' \to Z_{\psi'} \psi'$, with the
assumptions $Z_1=Z_\lambda Z_{\psi'}^2$, $Z_2=Z_{\psi'} Z_\psi$,
$Z_3=Z_{\psi'} Z_\lambda Z_\psi$, $Z_4=Z_{\psi'} Z_\lambda Z_\tau
Z_\psi$, $Z_5=Z_{\psi'} Z_\lambda Z_g Z_\psi^3$.

Important extra information about renormalization constants could
be obtained from consideration of the static (simultaneous)
Green's functions of the theory (\ref{initAction}). It could be
shown that such functions coincide with renormalized Green's
functions of $\varphi^4$ theory with static action
\begin{equation}\label{Sstat}
S_R^{\rm stat}(\Phi) = Z_3 \psi
\partial^2 \psi/2 -Z_4 \tau \psi^2/2 -Z_5 g \mu^{\eps} \psi^4/24,
\end{equation}
where static renormalization constants $Z_3$, $Z_4$ and $Z_5$ in
the MS scheme coincide with the respective dynamical ones in
(\ref{renormAction}) so the renormalization constants of the field
$\psi$ and parameters $g$ and $\tau$ are equal to the static
ones~\cite{V}. Moreover the requirement that the static and the
dynamical theories agree leads to relation
\begin{equation}\label{1}
 Z_{\psi'}Z_{\lambda}=Z_{\psi}\,,
\end{equation}
from which it follows
\begin{equation}
 Z_1=Z_2=Z_{\lambda}^{-1}Z_{\psi}^2\,,\quad Z_3=Z_{\psi}^2\,,
 \quad Z_4=Z_{\tau}Z_{\psi}^2\,,\quad Z_5=Z_gZ_{\psi}^4\,.
\end{equation}

The only new renormalization constant in addition to the static
theory is $Z_{\lambda}$ that can be found from the 1-irreducible
functions $\langle\psi'\psi'\rangle$ or
$\langle\psi'\partial_t\psi\rangle$. At present the static
renormalization constants are known up to 5-loop accuracy (fifth
order of the perturbation theory). So in the MS scheme one can use
coordinate of the fixed point $g^*$ (root of $\beta$-function
$\beta(g^*)=0$) calculated in the static model (\ref{Sstat}). In
the following in place of $g$ we will use for convenience the
charge $u=g{S_d}/{(2\pi)^d}$, where
$S_d={2\pi^{d/2}}/{\Gamma(d/2)}$ is the area of the unit sphere in
$d$ dimensions. The value of the charge at the fixed point $u^*$
is given in~\cite{V}. With the accuracy required for the following
calculation we have
\begin{equation}
u^*
 = \frac{2}{3 k_1} \eps + \frac{34}{81} \frac{k_0}{k_1^3} \eps^2 +
 \frac{2}{ 27 k_1^5 } \left( \frac{ -33n^3+110n^2+1760n+4544 }{5832 }
 - 4 \zeta(3) k_1 k_4 \right) \eps^3+O(\eps^4).\label{u}
\end{equation}
From now on we use the notation:
\begin{multline}
 k_0=(3n+14)/17,\ k_1=(n+8)/9,\ k_2=(n^2+6n+20)/27,\\
 k_3=(n+2)/3,\ k_4=(5n+22)/27,
\end{multline}
all $k_i$ are normalized so that $k_i=1$ at $n=1$.

\section{Calculation of the dynamical critical exponent}
The dynamical critical exponent $z$ is expressed by the relation
$z=2+\gamma^*-\eta$ by means of the Fisher critical exponent
$\eta$ and the value $\gamma^*\equiv \gamma(u^*)$ of RG-function
$\gamma(u)=\beta(u)\partial_u \ln Z_2$ at the fixed point
$u^*$~\cite{V}. Traditionally $z$ is written in the form
$z=2+R\eta$ with $R\equiv \gamma^*/\eta-1$.  The quantity $R$ is
convenient because the first two terms of its $\eps$-expansion
being independent of $n$ (the number of field components). Indeed
the leading terms of the $\eps$-expansion of the Fisher exponent
are determined by the series~\cite{V}
\begin{equation}
\eta = k_3 \frac{{u^*}^2}{24} \left[ 1 + a_1 k_1 u^* + ( a_2 k_2 +
a_3 k_3 + a_4 k_4 ) {u^*}^2 \right]+O({u^*}^5),\label{eta}
\end{equation}
where
\begin{equation}
 a_1=-3/8,\qquad a_2=-15/64,\qquad a_3=-5/32,\qquad
a_4=45/32,\label{a}
\end{equation}
while the analogous expansion for the $\gamma^*$ is given by
\begin{equation}
\gamma^* = k_3 h \frac{u^{*2}}{24} \left[ 1 + c_1 k_1 u^* + ( c_2
k_2 + c_3 k_3 + c_4 k_4 ) {u}^{*2} \right]+O({u^*}^5)
\label{gamma},
\end{equation}
with some constants $c_i$, $h$. The quantities (\ref{eta}) and
(\ref{gamma}) depend on $n$ only via $k_i$. Calculating the
quotient $\gamma^*/\eta$ from these relations and using (\ref{u})
one finds
\begin{equation}
  \frac{\gamma^*}{\eta} = h \left\{1 +
  \frac{2}{3}(c_1-a_1)\eps+\left[ \frac{4}{9}a_1(a_1-c_1) +
  \frac{4}{3}(c_2-a_2)+\frac{nb_1+b_0}{(n+8)^2}\right]\eps^2 + O(\eps^3)
  \right\},
  \label{gamma/eta}
\end{equation}
 where
\begin{gather}\label{b}\nonumber
  b_0= 28(c_1-a_1) - \frac{176}{3}(c_2-a_2) + 24(c_3-a_3) +
  \frac{88}{3}(c_4-a_4),\\
  b_1=6(c_1-a_1) - \frac{40}{3}(c_2-a_2) + 12(c_3-a_3) +
  \frac{20}{3}(c_4-a_4).
\end{gather}
It is seen from these relations that first two terms of the
$\eps$-expansion of $R$ indeed do not depend on $n$. Evaluation of
the renormalization constant $Z_2$ in 4-loop approximation gives
the following values of the coefficients in (\ref{gamma}):
\begin{equation}
 h=1.72609,\
 c_1=-0.4939(4),\ c_2=-0.2512(0),\ c_3=-0.1699(4),\
 c_4=1.806(3).
 \label{hc}
\end{equation}
The first calculation of $h$ was performed in~\cite{1/N} (one
2-loop graph):
\begin{equation}
h=6\ln (4/3).
 \label{h}
\end{equation}
The constant $c_1$ was calculated in~\cite{3loop1} and corrected
in~\cite{3loop} (three 3-loop graphs):
\begin{equation}
c_1=\frac{\pi^2/8-F(1/4)}{\ln (4/3)}-\frac{3}{4}+\frac{13}{8}\ln
4-\frac{21}{8}\ln 3,
 \label{c1}
\end{equation}
where $F(x)=\int_x^1 dt\ln t/(t-1)$ is dilogarithm (Spence's
function). The values of the constants $h$ and $c_1$ obtained in
(\ref{hc}) agree with the analytic expressions from (\ref{h}) and
(\ref{c1}).

In regard to the constants we pioneer in calculating of, namely,
$c_2$, $c_3$ and $c_4$ (twenty five 4-loop graphs). It turns out
that one can verify the value of $c_2$ using the asymptotic
expression for the constant $R$ in the limit
$n\to\infty$~\cite{1/N}:
\begin{equation}
R=\frac{4}{\eps}\left\{\frac{(4-\eps)\Gamma^2(1-\eps/2)}
{8\Gamma(2-\eps)\int_0^{1/2}dx[x(2-x)]^{-\eps/2}}-1\right\}+O(1/n)
\label{R}.
\end{equation}
Calculating of the $\eps$-expansion in the rhs of (\ref{R}) gives
\begin{equation}
  R=[6\ln(4/3)-1](1-0.188483 \eps-0.0999529 \eps^2)+O(\eps^3), \quad
  n\to\infty
  \label{Rinf}.
\end{equation}
On the other hand from (\ref{gamma/eta}) we obtain
\begin{equation}
  R=\frac{\gamma^*}{\eta}-1 = h \left\{ 1 + \frac{2}{3}(c_1-a_1)\eps
  + \frac{4}{9}[ a_1(a_1-c_1) + 3(c_2-a_2)] \eps^2
  \right\}-1+O(\eps^3),\quad n\to\infty \label{Rinf1}.
\end{equation}
After the substitution of the values from (\ref{a}) and (\ref{hc})
into (\ref{Rinf1}) the agreement between the coefficients of the
$\eps$-expansions in (\ref{Rinf}) and (\ref{Rinf1}) is
established.

The final expression of the constant $R$ is obtained by the
substitution of the quantities (\ref{hc}) to (\ref{b}) and
(\ref{gamma/eta}):
\begin{equation}
R=[6\ln(4/3)-1]\left(1-0.1884(9)\mult
\eps+\left[\frac{4.78(6)\mult n+21.5(4)}{(n+8)^2}-0.100(4)\right]
\eps^2\right)+O(\eps^3)
 \label{R4}.
\end{equation}
The final expression for the dynamical critical exponent
$z=2+R\eta$ for $n=1$ results from (\ref{1}), (\ref{eta}) and
(\ref{R4}):
\begin{equation}
\left. z \right|_{n=1} = 2 + 0.0134462 \mult \eps^2 + 0.011036(2)
\mult \eps^3 - 0.00558(5) \mult \eps^4+O(\eps^5). \label{VDZ}
\end{equation}

\section{Discussion}
The method that we used for extracting the residues of graphs at
$\eps=0$ and for the numerical calculating of the renormalization
constants, turned out to be highly effective for the RG-analysis
of the dynamical model $A$ in the 4-loop approximation.
Correctness of the obtained results was confirmed by the
cancellation of the poles at $\eps=0$ in the RG function
$\gamma(u)=\beta(u)\partial_u \ln Z_2$ and by comparison with the
results known in two and three loops and with the leading order of
the $1/n$-expansion. The main difficulty in the approach we used
was achieving of high precision in numerical calculations of
multiple integrals (in our case maximum multiplicity equals to
nine and the achieved precision is of order $0.1\%$.) The proposed
method can be used for multiloop calculations of the
renormalization constants for both dynamical and statical
problems. In particular, we are planning 6-loop calculation in the
static model (\ref{Sstat}).

It is obvious from  (\ref{VDZ}) that the 4-loop correction to the
critical exponent $z$ is relatively small for the real value of
$\eps=1$ (i.e. $d=3$) we obtain $z=2.0188(9)$.

\section*{Acknowledgements}
The authors thank N.V.~Antonov for discussions. The work was
supported by Russian Foundation for Basic Research (grant No
08-02-00125a).

\end{document}